\documentclass[conference]{IEEEtran}
\IEEEoverridecommandlockouts
\usepackage{blindtext, graphicx, geometry}
\usepackage{multirow, array, mathtools}

\usepackage{mathtools}
\usepackage{cite}
\usepackage{float}
\usepackage{multicol}
\usepackage{multirow}
\usepackage{hyperref}
\usepackage{graphicx}
\usepackage{amsfonts}
\usepackage{amsmath}
\usepackage{color}
\usepackage{epstopdf}
\usepackage{subcaption}
\usepackage{mathtools, nccmath}
\usepackage{multirow}
\usepackage{bbm}
\usepackage{marginnote}
\usepackage{tikz}
\usepackage{bm}
\usetikzlibrary{arrows,decorations.pathmorphing,backgrounds,fit,positioning,shapes.symbols,chains, shapes,snakes}
\usepackage{eurosym}
\usepackage{comment}

\usepackage[font={small}]{caption}

\ifCLASSINFOpdf
\else
\fi

\newenvironment{ldescription}[1]
  {\begin{list}{}%
   {\renewcommand\makelabel[1]{##1\hfill}%
   \settowidth\labelwidth{\makelabel{#1}}%
   \setlength\leftmargin{\labelwidth}
   \addtolength\leftmargin{\labelsep}}}
  {\end{list}}

\begin{document}

\title{IoT-enabled Distributed Cyber-attacks on Transmission and Distribution Grids.}

\author{ \parbox{5 in}{\centering Yury Dvorkin, \textit{IEEE}, \textit{Member},  and Siddharth Garg, \textit{IEEE}, \textit{Member} \\
Department of Electrical \& Computer Engineering \\
Tandon School of Engineering \\
New York University \\
         {\tt\small $\{$dvorkin, sg175$\}$@nyu.edu}}
}

\maketitle

\begin{abstract}
The Internet of things (IoT) will make it possible to interconnect and simultaneously control distributed electrical loads. Various technical and regulatory concerns have been raised that IoT-operated loads are being deployed  without appropriately considering and systematically addressing potential cyber-security challenges.  Hence, one can envision a hypothetical scenario when an ensemble of IoT-controlled loads can be hacked with malicious intentions of compromising operations of the electrical grid. Under this scenario, the attacker would use  geographically distributed IoT-controlled loads to alternate their net power injections into the electrical grid in such a way that may disrupt normal grid operations. 

 This paper presents a modeling framework to analyze grid impacts of distributed cyber-attacks on IoT-controlled  loads. This framework is used to demonstrate how a hypothetical distributed cyber-attack propagates from the distribution electrical grid, where  IoT-controlled  loads are expected to be installed, to the transmission electrical grid. The techno-economic interactions between the distribution and transmission electrical grids are accounted for by means of bilevel optimization. The case study is carried out on the modified versions of the 3-area IEEE Reliability Test System (RTS) and the IEEE 13-bus distribution feeder. Our numerical results demonstrate that the severity of such attacks depends on  the penetration level of IoT-controlled loads and the strategy of the attacker.  

\end{abstract}

\vspace{10pt}

%
\IEEEpeerreviewmaketitle

\section*{Nomenclature}

\subsection{Sets and Indices}
\begin{ldescription}{$xxxxxxxx$}
\item [$\mathcal{B}^{\text{T/D}}$] Set of buses in the transmission/distribution grid, indexed by $b$.
\item [$\mathcal{B}^{\text{D, $\downarrow$}}_{b}$] Set of buses in the distribution grid located downstream of bus $b$.
\item [$\mathcal{B}^{\text{+}}$] Set of buses in the transmission grid that are root buses for distribution grids, indexed by $b$.
\item [$\mathcal{I}^{\text{T/D}} (\mathcal{I}^{\text{T/D}}_b)$] Set of generation resources located in the transmission/distribution grid (located at bus $b$), indexed by $i$.
\item [$\mathcal{L}^{\text{T/D}}$] Set of lines in the transmission/distribution grid, indexed by $l$.
\item [$\mathcal{V}^{\text{A}}$] Set of decision variables of the attacker.
\item [$\mathcal{V}^{\text{T/D}}$] Set of decision variables made for the transmission/distribution grid.
\item [$\mathcal{V}^{\text{TD}}$] Set of dual decision variables for the transmission grid.
\item [$b_0$] Index of the root bus in the distribution grid.
\item [$o(l)/r(l)$] Indices of  sending/receiving buses of line $l$.
\end{ldescription}
\subsection{Parameters}
\begin{ldescription}{$xxxxxxxx$}
\item [$B_l$] Susceptance of  line $l$.
\item [$C^{\text{B}}_{b}$] Price bid of demand at bus $b$. 
\item [$C^{\text{O}}_{i}$] Price offer of generation resource $i$. 
\item [$\overline{F}_l$] Active power flow limit in line $l$.
\item [$\overline{G}^{\text{p}}_i/\underline{G}^{\text{p}}_i$] Upper/lower limits on the active power output of generator $i$.
\item [$\overline{G}^{\text{q}}_i/\underline{G}^{\text{q}}_i$] Upper/lower limits on the reactive power output of generator $i$.
\item [$G_l$] Conductance of line $l$.
\item [$L^{\text{p/q}}_b$] Active/reactive power load at bus $b$.
\item [$\overline{P}_b$] Upper limit on the active power transfer between the transmission and distribution grids at bus $b$ of the transmission grid.
\item [$R_l$] Resistance of line $l$.
\item [$T$] Tariff in the distribution grid.
\item [$\overline{S}_l$] Apparent power flow limit in line $l$.
\item [$\overline{V}_i/\underline{V}_i$] Current magnitude squared in line $l$ of the distribution grid.
\item [$X_l$] Reactance of line $l$.
\item [$\Gamma$] Attack severity parameter.
\item [$\overline{\Delta}^{\text{p}}, \underline{\Delta}^{\text{p}}$] Upper/lower limits on the net active power injection of IoT-controlled loads.
\item [$\overline{\Delta}^{\text{q}}, \underline{\Delta}^{\text{q}}$] Upper/lower limits on the net reactive power injection of IoT-controlled loads.
\item [$\omega_{lb}$] Power transfer distribution factor (PTDF) of line $l$ from injection at bus $b$. 
\end{ldescription}

\subsection{Variables}
\begin{ldescription}{$xxxxxxx$}
\item [$a_l$] Current magnitude squared in line $l$ of the distribution grid.

\item [$f_l^{\text{p/q}}$] Active/reactive power flow in line $l$.
\item [$g_i^{\text{p/q}}$] Active/reactive power output of generation resource $i$.
\item [$o^{\text{A,T/D}}$] Objective function of the attacker in the transmission/distribution grid. 
\item [$o^{\text{T/D}}$] Objective function of transmission/distribution grid.
\item [$o^{\text{TD}}$] Dual objective function of the transmission grid
\item [$p_b^{\text{B/O}}$] Active power bid/offered by the distribution utility at bus $b$ the wholesale electricity market.
\item [$v_b$] Voltage magnitude squared at bus $b$.
\item [$\Delta f_l^{\text{p/q}}$] Active/reactive power flow adjustment in line $l$ due to attacker's injection $\Delta l_b^{\text{p/q}}$ .
\item [$\Delta l^{\text{p/q}}$] Attacker's modification of the net active/reactive power injection of IoT-controlled loads at bus $b$. 
\item [$\theta_b$] Voltage angle at bus $b$.
\end{ldescription}

\section{Introduction}

Large-scale deployment of internet-based devices, at both the customer- and grid-ends, expose electrical grids to an imminent threat of cyber-attacks. These threats are much alike to other internet-connected services.
Previous studies have primarily focused on cyber-attacks 
that directly compromise 
power utilities, large generation companies or individual bulk generators~\cite{rosas2007topological,wang2011robustness}. 
On the other hand, \textit{distributed cyber-attacks} launched from an ensemble of compromised IoT-controlled loads  can be equally pernicious. This paper aims to analyze the 
threats posed by such distributed cyber-attacks. 


The penetration level of IoT-controlled devices in homes, offices and businesses  
that are directly connected to the electrical grid and can be 
compromised remotely has been constantly increasing. 
Compromised IoT devices have already been used to 
launch devastating distributed denial of service (DDoS) attacks on web services.
For instance, a recent attack on the Dyn, a domain name service (DNS) provider, caused major internet websites, e.g. Twitter,
to be unavailable for hours~\cite{dyn}. Using specialized malware, attackers captured myriads of IoT-controlled devices (using their default passwords) and routed traffic from these captured devices  to the Dyn. The distributed nature of this attack enabled the attackers to circumvent standard defense mechanisms (e.g., those that detect a large volume of traffic from a single IP address).

The scale and impact of the Dyn attack presages similar attacks 
on the electrical grid. Specifically, 
an attacker can potentially be able to
modify the power consumption
of compromised IoT-controlled loads  
to maliciously cause load shedding, reduce security margins, or even trigger a cascading failure. 
As opposed to conventional cyber-attacks on the 
grid that target a relatively small number of power utilities, large generation companies or individual bulk generators~\cite{rosas2007topological,wang2011robustness}, the distributed cyber-attacks considered in this paper
target a large number 
of small IoT-connected loads. Even though these loads can be small-scale and geographically distributed, potentially across multiple 
distribution grids, a single attacker is technically capable of aggregating these loads and modifying their power consumption
in a co-ordinated, centralized manner.

\begin{figure}[t]
    \centering
    \includegraphics[width=0.7\columnwidth]{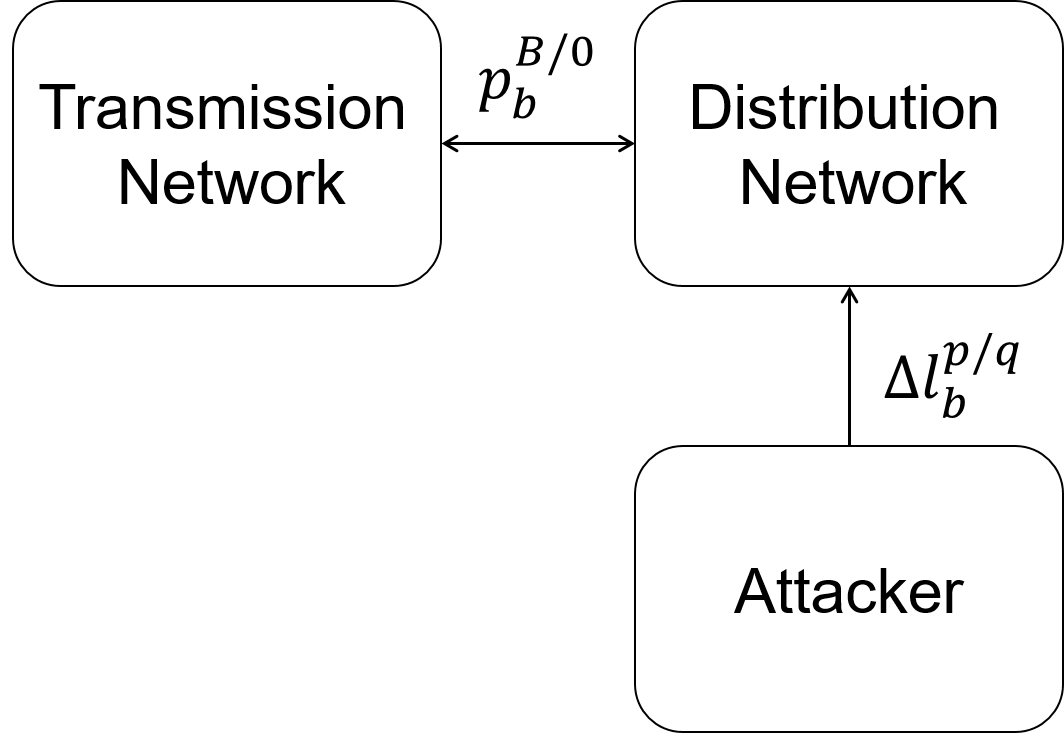}
    \caption{Interactions between the attacker and electrical grid.}
    \label{fig:overview}
\end{figure}

Figure~\ref{fig:overview} shows the interactions between the electrical grid and the attacker modeled in this paper. The attacker devises the 
optimal attack strategy to maximize damage to 
the electrical grid by modifying the active and reactive
power consumption of IoT-controlled loads by $\Delta l^{\text{p}}_b$ and  $\Delta l^{\text{q}}_b$, respectively. 
These modifications change the power flow in distribution lines, which in turn change the power flow in transmission lines. 
The interactions between the distribution and 
transmission grids are modeled by means of 
bi-level optimization. 
This paper models two attack strategies: naive and insidious. 
The naive strategy assumes that the attacker cannot forecast and circumvent the 
effect of built-in grid protection mechanisms, e.g. circuit breakers, to maximize  damage, while the insidious strategy accounts for these effects.  Based on the proposed framework, the case study demonstrates potential grid impacts of such distributed cyber-attacks. 

\subsection{Related Work}

Previous studies on cyber-attacks on the electrical grid have typically 
focused on centralized attacks in which an attacker gains access to 
the supervisory control and data acquisition (SCADA) system. For example, this attack mechanism led to the famous stuxnet attack~\cite{langner2011stuxnet}. Distributed cyber-attacks on the electrical grid have been considered
primarily in the context of false data injection attacks. For example, this attack mechanism is reported in~\cite{liu2011false,kosut2011malicious}, where  the attacker  hijacks and falsely reports
phasor measurement unit (PMU) readings. That is, the attacker attempts to maliciously 
change 
PMU readings  and trick the grid operator into making scheduling and dispatch decisions that hamper reliability. 

A cyber-attack on an relatively large aggregator of distributed IoT-controlled loads is modeled in \cite{mohsenian2011distributed}. However, \cite{mohsenian2011distributed} considers the transmission grid only and does not account for the interactions between the transmission and distribution grids, where IoT-controlled loads are expected to be deployed. Compared to \cite{mohsenian2011distributed}, this paper studies cyber-attacks on IoT-controlled loads located in the distribution grid. 


\subsection{Paper Contributions}

This paper makes the following contributions:
\begin{enumerate}

\item We describe a bilevel optimization model that makes it possible to operate the transmission and distribution electrical grid coordinately. 

\item We devise the optimal attack strategy for a naive and insidious attacker.

\item Using the bi-level optimization model and attack strategy described above, we analyze the impacts of distributed cyber-attacks on the distribution and transmission electrical grid. 

\end{enumerate}

The rest of this paper is organized as follows. Section~\ref{sec:grid_steady} describes the model that makes it possible to operate the transmission and distribution grids coordinately. Section~\ref{sec:attack} describes the model of the attacker. Section~\ref{sec:grid_attack} describes the model for analyzing the transmission and distribution grid operations under the attack.  Section~\ref{sec:case_study} describes the case study and Section~\ref{sec:conclusion} concludes the paper.

\section{Steady-State Grid Operations} \label{sec:grid_steady}
This section describes the models used for the coordinated operation of the distribution and transmission electrical grids. 

\subsection{Assumptions}
The transmission and distribution grid models are developed under the following assumptions:
\begin{enumerate}
\item The proposed transmission and distribution models are formulated for a single time period, i.e. inter-temporal and mixed-integer constraints are neglected.
\item The transmission grid is operated by the transmission system operator (TSO) using a lossless dc approximation of ac power flow equations that can accommodate a meshed network topology.  The TSO is responsible for clearing the whole-sale electricity market, which is assumed to be perfectly competitive.
\item The distribution grid  has a tree network topology and is operated by the distribution system operator (DSO) using an SOCP relaxation of ac power flow equations, i.e. it explicitly accounts for active power and reactive power flows, as well as for voltage limits.  The DSO is responsible for scheduling generation resources located in the distribution grid, as well as for purchasing and selling power in the electricity market. 
\item The transmission and distribution grids are connected via the single bus. This interface is limited to the active power exchange of a finite capacity, i.e.  the distribution grid has sufficient internal resources to compensate its reactive power needs. 
\end{enumerate}

\subsection{Distribution Grid}
\begin{flalign}
& \max_{\mathcal{V}^{\text{D}}} o^{\text{D}}\!:= \! \bigg[\!  \sum_{b \in \mathcal{B}^{\text{D}}} L^{\text{p}}_b T - \! \sum_{i \in \mathcal{I}^{\text{U}}} \! C^{\text{O}}_i  g_i^\text{p}  + \lambda_{b_0}  (p_{b_0}^{\text{O}} -   p_{b_0}^{\text{B}})\! \bigg] \label{dso_obj} \\
& g^{\text{p}}_i \leq \overline{G}^{\text{p}}_i, \quad \forall i \in \mathcal{I}^{\text{D}} \label{dso_eq1} \\
& g^{\text{p}}_i \geq \underline{G}^{\text{p}}_i, \quad \forall i \in \mathcal{I}^{\text{D}}\label{dso_eq2} \\
& g^{\text{q}}_i \leq \overline{G}^{\text{q}}_i, \quad \forall i \in \mathcal{I}^{\text{D}}\label{dso_eq3} \\
& g^{\text{q}}_i \geq \underline{G}^{\text{q}}_i, \quad \forall i \in \mathcal{I}^{\text{D}} \label{dso_eq4} \\
& (f^{\text{p}}_l)^2 + (f^{\text{q}}_l)^2 \leq \overline{S}_l^2, \quad \forall l \in \mathcal{L}^{\text{D}}  \label{dso_eq5} \\
& (f^{\text{p}}_l - a_l R_l)^2 +  (f^{\text{q}}_l - a_l X_l)^2  \leq \overline{S}_l^2, \quad \forall l \in \mathcal{L}^{\text{D}}  \label{dso_eq6} \\
& v_{r(l)} - 2 (R_l f^{\text{p}}_l + X_l f^{\text{q}}_l) + a_l (R_i^2 + X_l^2) =v_{o(l)}, \quad \forall l \in \mathcal{L}^{\text{D}} \label{dso_eq7} \\
& \big[(f^{\text{p}}_l)^2 + (f^{\text{q}}_l)^2\big]\frac{1}{a_l} \leq v_{o(l)}, \quad \forall l \in \mathcal{L}^{\text{D}}  \label{dso_eq8} \\
& f_{l|o(l)=b}^{\text{p}} -\!\! \sum_{l|r(l)=b  } \!\!(f_l^{\text{p}} - a_l R_l) - \sum_{i \in \mathcal{I}_b^{\text{D}}}g_i^{\text{p}} + L_b^{\text{p}} + v_b  G_{l|o(l)=b} \nonumber \\ &  \hspace{4.2cm} =0,  \quad  \forall b \in \mathcal{B}^{\text{D}} \backslash \big\{ b_0 \big\}\label{dso_eq9} \\
& f_{l|o(l)=b}^{\text{q}} -\!\! \sum_{l|r(l)=b  } \!\!(f_l^{\text{q}} - a_l X_l) - \sum_{i \in \mathcal{I}_b^{\text{D}}}g_i^{\text{q}} + L_b^{\text{q}} - v_b  B_{l|o(l)=b} \nonumber \\ &  \hspace{4.2cm}  =0,  \quad\forall b \in \mathcal{B}^{\text{D}} \backslash \big\{ b_0 \big\} \label{dso_eq10} \\
&  \!\!- \!\! \sum_{l|r(l)=b_0  } \!\!\!\!(f_l^{\text{p}} - a_l R_l) \!- \!p^{O}_{b_0} + p^{B}_{b_0} + v_{b_0}  G_{l|o(l)=b_0} =0 \label{dso_eq11} \\
&  \!\!- \!\!\sum_{l|r(l)=b_0  } \!\!(f_l^{\text{q}} - a_l X_l) - v_{b_0}  G_{l|o(l)=b_0}  =0 \label{dso_eq12} \\
&  p^{O}_{b0} \leq \overline{P}_{b0} \label{dso_eq13}
\end{flalign}
\begin{flalign}
&  p^{B}_{b0} \leq \overline{P}_{b0} \label{dso_eq14}\\
& v_b \leq \overline{V}_b, \quad \forall b \in \mathcal{B}^{\text{D}} \label{dso_eq15} \\
& v_b \geq \underline{V}_b, \quad \forall b \in \mathcal{B}^{\text{D}} \label{dso_eq16} \\
& \mathcal{V}^{\text{D}} = \big\{ a_l,  g^{\text{p}}_i, g^{\text{q}}_i, v_b\geq0; f^{\text{p}}_l, f^{\text{q}}_l\text{: free}  \big\} \label{dso_eq17}
\end{flalign}
Eq.~\eqref{dso_obj} is the objective function of the DSO, which aims to maximize the gross profit that includes the following three terms: i) payments collected by the utility from its customers, ii) operating cost of generation resources located in the distribution grid, iii) cost of transactions that DSO performs in the wholesale electricity markets. If $p^{\text{O}}_{b0} > 0$, i.e. the DSO sells surplus electricity in the wholesale electricity market, the third term of the objective function is positive. Alternatively, if  $p^{\text{B}}_{b0} < 0$, i.e. the DSO purchases electricity in the wholesale electricity market, the third term is negative. Note that due to the market regulation, there is a need to enforce complementarity condition on  decisions $p^{\text{B}}_{b0}$ and $p^{\text{O}}_{b0}$ that ensures that the DSO cannot simultaneously sell and purchase electricity in the wholesale market, i.e. $p^{\text{B}}_{b0} p^{\text{O}}_{b0} = 0$. Eq.~\eqref{dso_eq1}-\eqref{dso_eq4} enforce the minimum and maximum limits on the active and reactive power output of generation resources located in the distribution grid. Eq.~\eqref{dso_eq5}-\eqref{dso_eq6} limit the apparent power flow in distribution lines, while eq.~\eqref{dso_eq7}-\eqref{dso_eq9} represent a relaxation of ac power flow equations, \cite{6507355}. The active and reactive power balance for each bus is enforced in eq.~\eqref{dso_eq9}-\eqref{dso_eq12}. Note that eq.~\eqref{dso_eq11}-\eqref{dso_eq12} model the power balance for the root bus of the distribution grid with no  generation resource or load connected to that bus. The active power exchange between the distribution and transmission grid is limited in eq.~\eqref{dso_eq13}-\eqref{dso_eq14}.  Eq.~\eqref{dso_eq15}-\eqref{dso_eq16} enforce the maximum and minimum limits on bus voltages. Eq.~\eqref{dso_eq17} declares all decisions variables optimized by the DSO.

\subsection{Transmission Grid}
\begin{flalign}
& \max_{\mathcal{V}^{\text{T}}} o^{\text{T}} := \nonumber \\ &  \hspace{0.5cm} \bigg[   \sum_{b \in \mathcal{B}^{\text{T}}} \!\! C^{\text{B}}_b D^{\text{p}}_b\!+\!\!\! \sum_{b \in \mathcal{B}^{\text{D}}}\!\! C_b^{\text{B}} p^{\text{B}}_b - \sum_{i \in \mathcal{I}^{\text{T}}} \!\! C^{\text{O}}_i  g^{\text{p}}_i\! -\!\! \sum_{i \in \mathcal{I}^{\text{U}}} \!\! C^{\text{O}}_b p^{\text{O}}_b \bigg] \label{tso_obj} \\
& \sum_{i \in \mathcal{I}_b} g^{\text{p}}_i   + \sum_{l|r(l)=b} f_l^{\text{p}} - \sum_{l|o(l)=b} f^{\text{p}}_l + p^{O}_b- p^{B}_b= 
 L^{\text{p}}_b
 :(\lambda_b),  \nonumber \\ 
&  \hspace{4.9cm} \quad  \forall b \in  \mathcal{B}^{\text{+}} \label{tso_eq1}  \\
& \sum_{i \in I_b} g^{\text{p}}_i   + \sum_{l|r(l)=b} f_l^{\text{p}} - \sum_{l|o(l)=b} f^{\text{p}}_l = 
 L^{\text{p}}_b
 :(\lambda_b),  \nonumber \\ 
&  \hspace{4.9cm} \quad  \forall b \in \mathcal{B}^{\text{T}} \backslash \big\{ \mathcal{B}^{\text{+}} \big\} \label{tso_eq1b}  \\
& g^{\text{p}}_i \leq \overline{G}^{\text{p}}_i  : (\overline{\alpha}_i),  \quad  \forall i \in \mathcal{I}^{\text{T}} \label{tso_eq2} \\
& g^{\text{p}}_i \geq \underline{G}^{\text{p}}_i  : (\underline{\alpha}_i),  \quad  \forall i \in \mathcal{I}^{\text{T}} \label{tso_eq3}\\
&  p^{O}_b \leq \overline{P}_b  :(\overline{\psi}_b ), \quad  \forall b \in \mathcal{B}^{\text{+}} \label{tso_eq4}\\
&  p^{B}_b \leq \overline{P}_b  :(\underline{\psi}_b), \quad  \forall b \in \mathcal{B}^{\text{+}} \label{tso_eq5}\\
& f^{\text{p}}_l = \frac{1}{X_l} ( \theta_{o(l)} - \theta_{r(l)}) :(\xi_l),  \quad  \forall l \in \mathcal{L}^{\text{T}} \label{tso_eq6} \\
& f^{\text{p}}_l \leq \overline{F}_l  : (\overline{\delta}_l), \quad \forall l \in \mathcal{L}^{\text{T}} \label{tso_eq7}\\
& f^{\text{p}}_l \geq -\overline{F}_l: (\underline{\delta}_l),   \quad  \forall l \in \mathcal{L}^{\text{T}} \label{tso_eq8}\\
& \mathcal{V}^{\text{T}} = \big\{ g^{\text{p}}_i, f^{\text{p}}_l, \theta_b\text{: free};  p^{O}_b, p^{B}_b\geq 0  \big\}.\label{tso_eq9}
\end{flalign}
Eq.~\eqref{tso_obj} is the objective function of the TSO, which aims to maximize the social welfare that includes the following terms: i) bids of the customers connected to transmission grid, ii) bids of the DSO, iii) offers of generators connected to the transmission grid, iv) offers of the DSO. The nodal active power balance is enforced in eq.~\eqref{tso_eq1} for every bus in the transmission grid connected to the distribution grid and in eq.~\eqref{tso_eq1b} for the remaining buses.   Eq.~\eqref{tso_eq2}-\eqref{tso_eq3} enforces the  maximum and minimum  limits on the active power output of the generation resources located in the transmission grid. Eq.~\eqref{tso_eq4}-\eqref{tso_eq5} limit the active power purchased or sold by the DSO. The active power flow in each line of the transmission grid is computed in eq.~\eqref{tso_eq6}  and eq.~\eqref{tso_eq7}-\eqref{tso_eq8} enforce the active power flow limits. Primal decision variables of eq.~\eqref{tso_obj}-\eqref{tso_eq8} are declared in \eqref{tso_eq9}, while dual variables are given in line with respective constraints after a colon.

\subsection{Coordinated DSO-TSO Model}
For the DSO model in eq.~\eqref{dso_obj}-\eqref{dso_eq17} and  the TSO model in eq.~\eqref{tso_obj}-\eqref{tso_eq9}, we formulate the following model to model their coordinated operation:
\begin{flalign}
& \max_{\mathcal{V}^{\text{D}}} o^{\text{D}} \label{bil_eq1}\\
& D (\mathcal{V}^{\text{D}} ) \leq 0 \label{bil_eq2} \\
& p_{b_0}^{\text{B}}, p_{b_0}^{\text{O}}, \lambda_{b_0} \in \text{arg } \max_{\mathcal{V}^{\text{T}}} o^{\text{T}}  \label{bil_eq3} \\
& \hspace{2.7cm} T (\mathcal{V}^{\text{T}} ) \leq 0,     \label{bil_eq4}
\end{flalign}
where $T(\cdot)$ and $D(\cdot)$ represent the constraints in eq.~\eqref{dso_eq1}-\eqref{dso_eq16} and \eqref{tso_eq1}-\eqref{tso_eq8}, respectively.

\section{Attacker's Model} \label{sec:attack}

\subsection{Assumptions}
The following assumptions are made on the attacker:
\begin{enumerate}
    \item The attacker has  perfect knowledge of all compromised IoT-controlled loads, their technical characteristics and placement in the distribution grid. From the attacker perspective, this assumption results in the  \textquoteleft best-case'  attack scenario. 
    \item  The attacker has perfect knowledge of the transmission and distribution topology, including the  power transfer distribution factors ($\omega_{lb}$). In some cases, this data can be obtained or estimated from open sources, e.g. \cite{6039476}. From the grid perspective, this assumption results in the \textquoteleft worst-case' attack scenario.    
\end{enumerate}

\subsection{Attacker's Model}
\begin{flalign}
& \max_{\mathcal{V}^{\text{A}}} \big[ (1-\Gamma) o^{\text{A,D}} + \Gamma o^{\text{A,T}} \big] \label{eq_att_1} \\
& \underline{\Delta}^{\text{p}}_b \leq \Delta l^{\text{p}}_b \leq \overline{\Delta}^{\text{p}}_b, \quad \forall b \in \mathcal{B}^{\text{U}} \label{eq_att_2} \\
& \underline{\Delta}^{\text{q}}_b \leq \Delta l^{\text{q}}_b \leq \overline{\Delta}^{\text{q}}_b, \quad \forall b \in \mathcal{B}^{\text{U}}  \label{eq_att_3} \\
& \mathcal{V}^{\text{A}} = \big\{ \Delta l^{\text{p}}_b, \Delta l^{\text{q}}_b:\text{ free} \big\} \label{eq_att_4} .
\end{flalign}
Eq.~\eqref{eq_att_1} is  the objective function of the attacker that jointly accounts for the adversary effects in the distribution ($o^{\text{A,D}}$) and transmission ($o^{\text{A,T}}$) grids. Parameter $\Gamma \in [0,1]$ is used by the attacker to vary the severity of  adversary impacts between the transmission and distribution grids. If $\Gamma = 0$, the attacker intends to affect the distribution grid only, whereas $\Gamma = 1$ indicates the intention to affect the transmission grid only.  Eq.~\eqref{eq_att_2}-\eqref{eq_att_3}  define the technical limits on the ability of the attacker to modify net active and reactive power injections of IoT-controlled loads, $\Delta l^{\text{p}}_b$ and $ \Delta l^{\text{q}}_b$, located in the distribution grid.  Eq.~\eqref{eq_att_4} declares decision variables of the attacker. 

Functions $o^{\text{A,D}}$ and $o^{\text{A,T}}$ can be formulated in multiple ways depending on the intentions of the attacker. In this paper, we assume that the only intention of the attacker is to cause line overload in the transmission and distribution grids by changing net power injections of IoT-controlled loads, $\Delta l^{\text{p}}_b$ and $\Delta l^{\text{q}}_b$,  located in the distribution grid.  The impact of $\Delta l^{\text{p}}_b$ and $\Delta l^{\text{q}}_b$ on power flows in a tree-structured distribution lines, can be computed as:
\begin{flalign}
& \Delta f^{\text{p}}_l = \sum_{b \in \mathcal{B}^{\text{D, $\downarrow$}}_{ r(l)}} \Delta l^{\text{p}}_b, \quad \forall l \in \mathcal{L}^{\text{D}} \label{attacker_dist1} \\ 
& \Delta f^{\text{q}}_l = \sum_{b \in \mathcal{B}^{\text{D, $\downarrow$}}_{r(l)}} \Delta l^{\text{q}}_b, \quad \forall l \in \mathcal{L}^{\text{D}}. \label{attacker_dist2}
\end{flalign}
Eq.~\eqref{attacker_dist1}-\eqref{attacker_dist2} compute the change in active and reactive power flows caused by altering IoT-controlled loads at the buses that are located downstream from the receiving bus of line $l$ ($r(l)$). Since the dc power flow approximation is assumed for the transmission grid, only the active power flows will be affected by the attacker:
\begin{flalign}
& \Delta f^{\text{p}}_l = \begin{cases} 
\sum_{b \in \mathcal{B}^{\text{+}}}\omega^{\text{T}}_{lb} \sum_{b \in \mathcal{B}^{\text{D}}} \Delta l_b^{\text{p}},   & \text{ if $\omega_{lb}^{\text{T}}>0$} ,\forall l \in \mathcal{L}^{\text{T}}  \\ 
-\sum_{b \in \mathcal{B}^{\text{+}}}\omega^{\text{T}}_{lb} \sum_{b \in \mathcal{B}^{\text{D}}} \Delta l_b^{\text{p}},   & \text{ if $\omega_{lb}^{\text{T}}<0$} ,\forall l \in \mathcal{L}^{\text{T}}  \\   
\end{cases} \label{attacker_trans}
\end{flalign}
Eq.~\eqref{attacker_trans} computes the active power flow change in transmission lines caused by $\Delta l^{\text{p}}_b$ based on power transfer distribution factors $\omega_{lb}$. Using Eq.~\eqref{attacker_dist1}-\eqref{attacker_trans}, functions  $o^{\text{A,U}}$ and $o^{\text{A,T}}$ can be defined as follows:
\begin{flalign} 
& o^{\text{A,D}} = \sum_{l \in \mathcal{L}^{D}}   (\Delta f^{\text{p}}_l)^2 +  (\Delta f^{\text{q}}_l)^2 \\
& o^{\text{A,T}} = \sum_{l \in \mathcal{L}^{T}} (\Delta f^{\text{p}}_l)^2
\end{flalign}

\section{Grid under the Attack} \label{sec:grid_attack} 
Given the attack described in eq.~\eqref{eq_att_1}-\eqref{eq_att_4}, parameters $D_b^{\text{p}}$ and $D_b^{\text{q}}$ are attain the following values:
\begin{flalign}
&  D_b^{\text{p}} \rightarrow D_b^{\text{p}} + \Delta l^{\text{p}}_b \label{eq_grid_at1}\\
&  D_b^{\text{q}} \rightarrow D_b^{\text{q}} + \Delta l^{\text{q}}_b \label{eq_grid_at2}
\end{flalign}
Using eq.~\eqref{eq_grid_at1}-\eqref{eq_grid_at2}, the post-attack values of power flows in the distribution and transmission grids can be obtained from the DSO-TSO model in eq.~\eqref{bil_eq1}-\eqref{bil_eq4}. This can be achieved if the model of the attacker and the coordinated DSO-TSO model are co-optimized. To enable this co-optimization, the duality based approach \cite{arroyo_vulnerability} is used to equivalently recast the bilevel DSO-TSO model in eq.~\eqref{bil_eq1}-\eqref{bil_eq4} as the following  single-level problem:
\begin{flalign}
& \max_{\mathcal{V}^{\text{D}} \cup \mathcal{V}^{\text{T}} \cup \mathcal{V}^{\text{TD}}} o^{\text{D}} \label{sle_eq1}\\
& D (\mathcal{V}^{\text{D}} ) \leq 0 \label{sle_eq2} \\
& T (\mathcal{V}^{\text{T}} ) \leq 0    \label{sle_eq3} \\
& T^{\text{TD}} (\mathcal{V}^{\text{TD}} ) \leq 0     \label{sle_eq4} \\
& o^{\text{T}}  =  o^{\text{TD}}   \label{sle_eq4}, 
\end{flalign}
where eq.~\eqref{sle_eq2}-\eqref{sle_eq3} list the primal constraints of the distribution and transmission grid models,  $T^{\text{TD}}(\cdot)$ in eq.~\eqref{sle_eq4} denotes the dual constraints of the transmission grid model and eq.~\eqref{sle_eq4} enforces the strong duality condition by equating the primal ($o^{\text{T}}$) and dual ($o^{\text{T}}$) objective functions of the transmission grid model. Note that $o^{\text{D}}$ in eq.~\eqref{sle_eq1} contains the bilinear product $\lambda_{b_0}  (p_{b_0}^{\text{O}} -   p_{b_0}^{\text{B}})$, which can be be equivalently replaced using complementarity slackness conditions of the transmission grid model with the following linear expression:
\begin{flalign}
\lambda_{b_0}  (p_{b_0}^{\text{O}} -   p_{b_0}^{\text{B}}) = C_b^{\text{O}} p_b^{\text{O}} - \overline{\psi}_b \overline{P}_b^{\text{O}}  + \underline{\psi}_b \underline{P}_b^{\text{B}} - C_b^{\text{B}} p_b^{\text{B}}. \label{lin_eq7} 
\end{flalign}

The linearized single-level equivalent of the DSO-TSO model in eq.~\eqref{bil_eq1}-\eqref{bil_eq4} and the model of the attacker in eq.~\eqref{eq_att_1}-\eqref{eq_att_4} constitute an equlibrium problem that can be solved using off-the-shelf solution technique. This description is omitted in this paper due to the pagination limit.


 \section{Case Study} \label{sec:case_study} 

\begin{figure}[b]
    \centering
    \includegraphics[width=0.9\columnwidth]{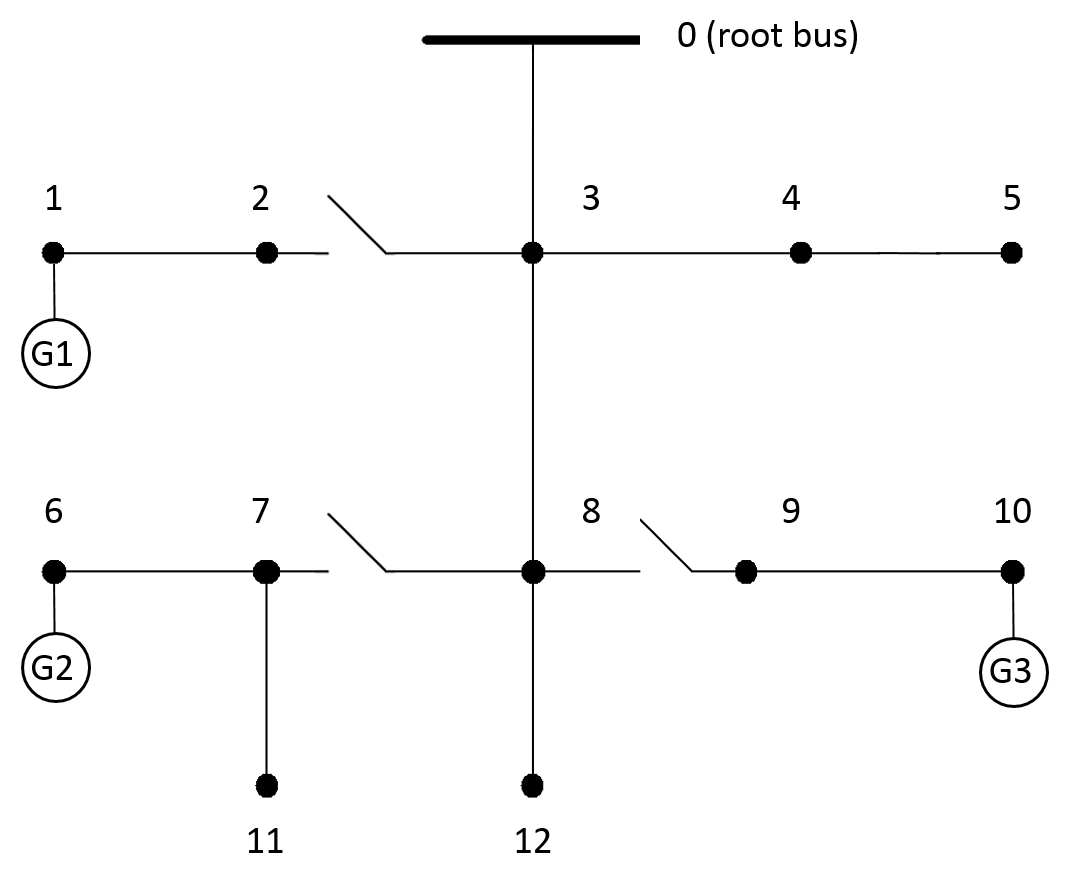}
    \caption{A schematic representation of the IEEE 13-bus test feeder \cite{Ieee_dist_feeder}. The total demand of 36 MW is equally distributed between buses \#2-5,-9, 11-12. Gas-fired generators with the maximum capacity of 5 MW are installed at buses \#1,6,10. The circuit breakers at branches \#2-3, 7-8 and 8-9 are closed during normal operations. }
    \label{fig:ieee_feeder}
\end{figure}

This case study uses the modified version of the 3-area IEEE Reliability Test System (RTS) \cite{pandzic_2014} for the transmission grid and the modified version of the IEEE 13-bus test feeder \cite{Ieee_dist_feeder} for the distribution grid. The transmission grid includes 96 conventional generators, 51 load buses, 120 transmission lines, and 19 wind farms. The total demand in the transmission grid is 8.9 GW and the total wind power output is set to 890 MW (10\% of the total demand). The transmission line ratings are reduced to 80\% of the original values to create congestion. The root bus of the distribution grid is connected to the transmission grid at bus \#102 of the 3-area IEEE Reliability Test System.  The distribution grid supplies the total active power demand of 36 MW and is populated with three 5 MW gas-fired generators with the incremental fuel cost of \$ 10/MWh, as displayed in Fig.~\ref{fig:ieee_feeder}. The power factor at each bus is kept at 0.9. The root bus of the distribution grid has an infinite supply of reactive power. The distribution line flow limits are adjusted to accommodate the apparent power flow to serve the active and reactive power demand at each bus  with a 10\% margin. It is assumed that the circuit breakers at both ends of each branch are set to automatically open if the apparent flow in that branch exceeds the rated value by 20\%. Thus, the power flow limit at branch 0-3 is 31.5 MVA and the circuit breaker at the root bus closes at the apparent flow equal to 34.4 MVA. Finally, we assume that the attacker can modify the power output of IoT-controlled loads  in their full operating range. 

To evaluate the socio-economic damage caused by the attacker we measure the energy not served, i.e. interrupted power supply (does not include line losses), and the cost it incurs defined as the product of the energy not served and the value of lost load. The value of lost load (VOLL) is set to \$10000/MW \cite{6847238}.

All simulations are carried out using Julia \cite{miles_julia2015} on a 2.9 GHz Intel Core i5 processor with 8GB of RAM.

\subsection{Normal operations}
During normal operations, three gas-fired generators G1-G3 output 5 MW each and the rest of the  active power demand (36$-$3$\times$5=21 MW) and active power line losses (1.3 MW) are consumed from the transmission grid leading to the apparent power flow in branch 0-3 of 28.7 MVA. 

\subsection{Attack on the distribution grid } \label{dist_grid}
If the attacker intends to limit the scope of the attack to the distribution grid, it derives the attacks strategy for $\Gamma =0$. Given the tree structure of the distribution grid, the attacker would aim to disconnect the distribution grid from the transmission grid, i.e. to open the circuit breaker at the root bus, and/or to isolate gas-fired generators by opening circuit breakers at branches 2-3, 7-8, 8-9.

The attacker can design two strategies: 
\begin{itemize}
\item[ i)] \textit{Naive strategy}: In this case the attacker is not aware of the settings on each circuit breaker and, therefore, cannot factor in the possibility of downstream disconnections that may in fact reduce the apparent power flow in branch 0-3.  

\item[ ii)] \textit{Insidious strategy}: The attacker accounts for the effect of downstream disconnections by  enforcing limits on $\Delta f_l^{\text{p}}$ and $\Delta f_l^{\text{q}}$ to avoid opening circuit breakers on branches  2-3, 7-8, 8-9, while maximizing the flow in branch 0-3.
\end{itemize}

Table~\ref{table_dist_attack_summary} summarizes the socio-economic impacts of the naive and insidious attack strategies for different penetration levels\footnote{The penetration level is defined as the ratio between the power of IoT-controlled loads at every bus to the total power demand at that bus.} of IoT-controlled loads.  As the penetration level remains low, the attacker has no effect on the distribution grid regardless of the attack strategy chosen. However, as the penetration level increases, both the energy not served and its cost gradually increases.   This trend indicates the need to develop a protection mechanism against such attacks for future grid operations that are likely to host more IoT-controlled loads. Furthermore, it is noteworthy the naive strategy leads to lower socio-economic impacts. This observation can be explained by the fact that downstream disconnections on branches 2-3, 7-8, 8-9 isolate some of the attacked IoT-controlled loads that reduces the flow in branch 0-3 and makes it more challenging for the attacker to open the circuit breaker at the root bus. The attacker succeeds to open the circuit breaker at the root bus and disconnect the transmission and distribution grids only under the insidious strategy with the 50\% penetration level of IoT-controlled loads. Note that in the latter case three gas-fired generators located in the distribution grid cover 15 MW of the total demand.

\begin{center}
\begin{table}[t]
 \captionsetup{justification=centering, labelsep=period, font=footnotesize, textfont=sc}
\caption{Socio-Economic Impacts on the Distribution Grid Measured in Terms of the Energy Not Served (ENS, MW) and Its Cost ($C^{\text{ENS}}$, \$) }
\begin{tabular}{ c |c |c | c |c |c |c }
\hline
\hline
\multirow{3}{*}{Strategy}  & \multicolumn{6}{c}{Penetration level of IoT-controlled loads} \\
\cline{2-7}
& \multicolumn{2}{c|}{10\%} & \multicolumn{2}{c|}{25\%} & \multicolumn{2}{c}{50\%} \\
\cline{2-7}
 & ENS & $C^{\text{ENS}}$  &   ENS & $C^{\text{ENS}}$ &  ENS & $C^{\text{ENS}}$ \\
\hline
Naive & 0& 0& 6.3 & 63,411& 9.1 & 90,623 \\
Insidious &0 & 0 &7.9 & 79,861& 21.0 &213,411\\
\hline
\end{tabular}
\label{table_dist_attack_summary}
\end{table}
\vspace{-15pt}
\end{center}
\vspace{-10pt}

\subsection{Attack on the transmission grid} 
\begin{figure}
    \centering
    \includegraphics[width=\columnwidth]{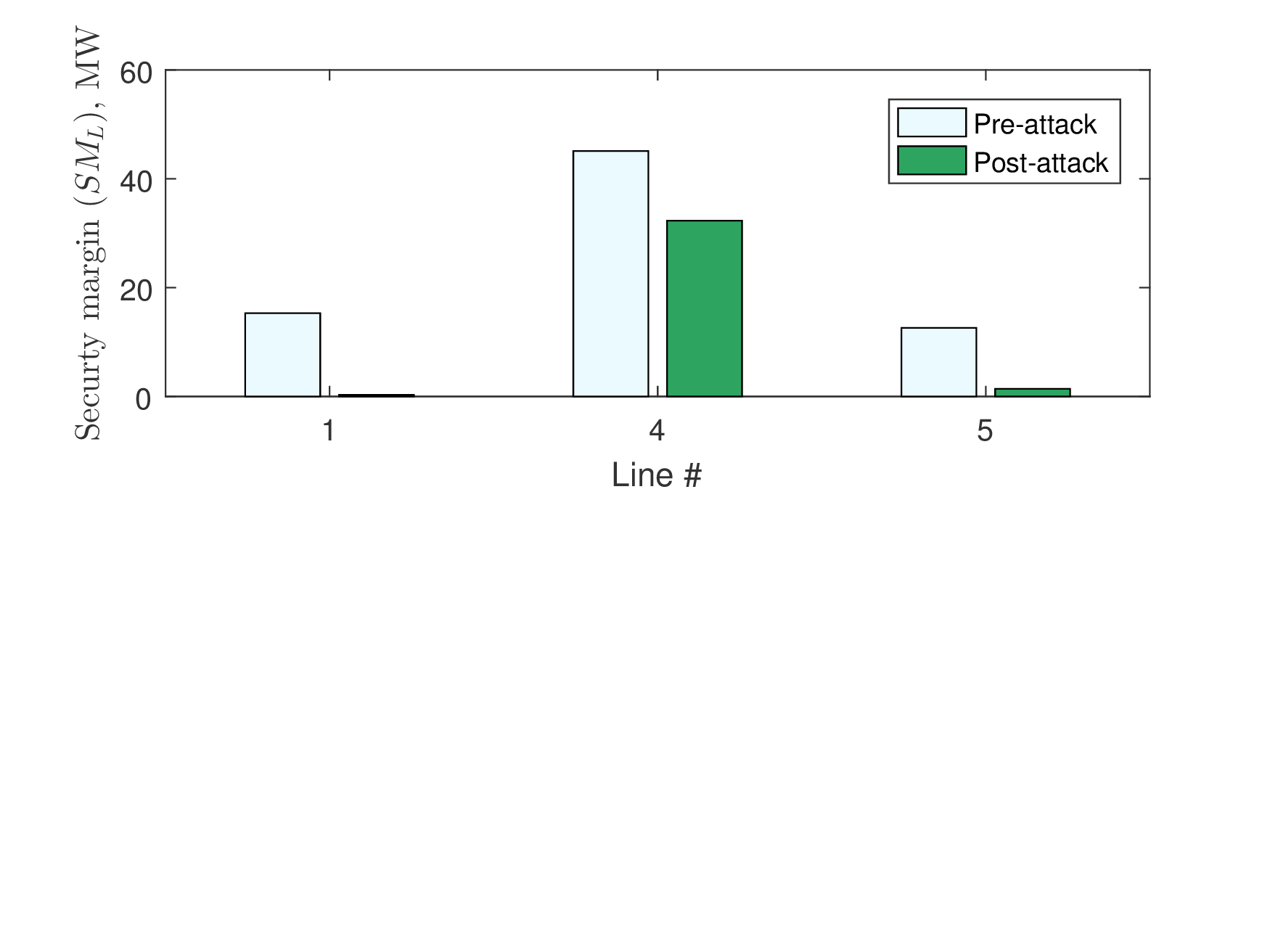}
    \vspace{-100pt}
    \caption{Comparison of pre- and post-attack security margins on transmission lines connected to bus \#102. }
    \vspace{-20pt}
    \label{fig:trans}
\end{figure}

This attack strategy is derived for $\Gamma =1$. In this case the attacker  aims to cause  damage to the transmission grid only and uses the distribution grid as a proxy. This attack is possible, if the attacker avoids opening the circuit breaker at the root bus of the distribution grid. Therefore, the attacker uses {insidious strategy} strategy from Section~\ref{dist_grid} amended with the setting of the circuit breaker at the root bus. Under the 10\% and 25\% penetration levels of the IoT-controlled loads in the distribution grid, the attacker has insignificant impacts on the transmission grid operations. However, as the penetration level increases to 50\%, the attacker seriously reduces security margins\footnote{The security margin ($SM_{l}$) for line $l$ is defined as $SM_{l} =\overline{F}_l - |f_l^{\text{p}}| $.} on transmission lines connected to bus \#102, as shown in Fig.~\ref{fig:trans}. Reduced security margins do not cause socio-economic impact per se but may trigger or contribute to line failures that in turn may cause cascading failures \cite{1519725}.

\section{Conclusion} \label{sec:conclusion}

This paper has described a modeling framework for analyzing the grid impacts of distributed attacks on IoT-controlled loads. The proposed model of grid operations accounts for both the transmission and distribution grid operations and makes it possible to analyze a propagation mechanism of such attacks in a bottom-up manner. The case study has demonstrated that the socio-economic impacts of such attacks on the distribution grid depends on the ability of the attacker to account for the effect of circuit breakers. Thus, the naive attack strategy without considering the effect of circuit breakers is shown to be less harmful than the insidious attack strategy. The ability of the attacker to noticeably affect transmission grid operations depends on a penetration level of IoT-controlled loads.  

The paper paves the way for several enhancements that will constitute our future work. The proposed modeling framework needs to be extended to account for the multi-period dynamics of power system operations and recourse actions under attack. This extension will make it possible to realistically evaluate the ability of the distribution and transmission grid to withstand such attacks via post-attack corrective actions and in presence of operational uncertainties, \cite{7268773}. Next, the attack propagation needs to be assessed in a probabilistic framework. Specifically, the likelihood of cascade failures in the transmission network needs to be evaluated.  From the perspective of the attacker, it is important to improve the modeling accuracy of individual IoT-controlled loads. In order to model the attack closer to a realistic environment, our future work will explore how publicly available grid sources can be used by the attacker to design more harmful attack strategies.

\bibliographystyle{IEEEtran}
\bibliography{ConferenceBib}

\begin{thebibliography}{10}
\providecommand{\url}[1]{#1}
\csname url@samestyle\endcsname
\providecommand{\newblock}{\relax}
\providecommand{\bibinfo}[2]{#2}
\providecommand{\BIBentrySTDinterwordspacing}{\spaceskip=0pt\relax}
\providecommand{\BIBentryALTinterwordstretchfactor}{4}
\providecommand{\BIBentryALTinterwordspacing}{\spaceskip=\fontdimen2\font plus
\BIBentryALTinterwordstretchfactor\fontdimen3\font minus
  \fontdimen4\font\relax}
\providecommand{\BIBforeignlanguage}[2]{{%
\expandafter\ifx\csname l@#1\endcsname\relax
\typeout{** WARNING: IEEEtran.bst: No hyphenation pattern has been}%
\typeout{** loaded for the language `#1'. Using the pattern for}%
\typeout{** the default language instead.}%
\else
\language=\csname l@#1\endcsname
\fi
#2}}
\providecommand{\BIBdecl}{\relax}
\BIBdecl

\bibitem{rosas2007topological}
M.~Rosas-Casals, S.~Valverde, and R.~V. Sol{\'e}, ``Topological vulnerability
  of the european power grid under errors and attacks,'' \emph{International
  Journal of Bifurcation and Chaos}, vol.~17, no.~07, pp. 2465--2475, 2007.

\bibitem{wang2011robustness}
J.-W. Wang and L.-L. Rong, ``Robustness of the western united states power grid
  under edge attack strategies due to cascading failures,'' \emph{Safety
  science}, vol.~49, no.~6, pp. 807--812, 2011.

\bibitem{dyn}
``Lessons from the dyn ddos attack,''
  \url{https://www.schneier.com/blog/archives/2016/11/lessons_from_th_5.html},
  accessed: 2017-04-12.

\bibitem{langner2011stuxnet}
R.~Langner, ``Stuxnet: Dissecting a cyberwarfare weapon,'' \emph{IEEE Security
  \& Privacy}, vol.~9, no.~3, pp. 49--51, 2011.

\bibitem{liu2011false}
Y.~Liu, P.~Ning, and M.~K. Reiter, ``False data injection attacks against state
  estimation in electric power grids,'' \emph{ACM Transactions on Information
  and System Security (TISSEC)}, vol.~14, no.~1, p.~13, 2011.

\bibitem{kosut2011malicious}
O.~Kosut, L.~Jia, R.~J. Thomas, and L.~Tong, ``Malicious data attacks on the
  smart grid,'' \emph{IEEE Trans. Sm. Gr.}, vol.~2, no.~4, pp. 645--658, 2011.

\bibitem{mohsenian2011distributed}
A.-H. Mohsenian-Rad and A.~Leon-Garcia, ``Distributed internet-based load
  altering attacks against smart power grids,'' \emph{IEEE Trans. Sm. Gr.},
  vol.~2, no.~4, pp. 667--674, 2011.

\bibitem{6507355}
M.~Farivar and S.~H. Low, ``Branch flow model: Relaxations and convexification
  -- part i,'' \emph{IEEE Transactions on Power Systems}, vol.~28, no.~3, pp.
  2554--2564, Aug 2013.

\bibitem{6039476}
J.~E. Price and J.~Goodin, ``Reduced network modeling of wecc as a market
  design prototype,'' in \emph{2011 IEEE Power and Energy Society General
  Meeting}, July 2011, pp. 1--6.

\bibitem{arroyo_vulnerability}
J.~M. Arroyo, ``Bilevel programming applied to power system vulnerability
  analysis under multiple contingencies,'' \emph{IET Gen., Trans. \& Dist.},
  vol.~4, no.~2, pp. 178--190, February 2010.

\bibitem{Ieee_dist_feeder}
``Distribution test feeders: 13 bus ieee test feeder,''
  \url{https://ewh.ieee.org/soc/pes/dsacom/testfeeders/}, 1992.

\bibitem{pandzic_2014}
H.~Pandzic, Y.~Wang, T.~Qiu, Y.~Dvorkin, and D.~S. Kirschen, ``Near-{Optimal}
  {Method} for {Siting} and {Sizing} of {Distributed} {Storage} in a
  {Transmission} {Network},'' \emph{IEEE Transactions on Power Systems},
  vol.~30, no.~5, pp. 2288--2300, Sep. 2015.

\bibitem{6847238}
Y.~Dvorkin, H.~Pandžić, M.~A. Ortega-Vazquez, and D.~S. Kirschen, ``A hybrid
  stochastic/interval approach to transmission-constrained unit commitment,''
  \emph{IEEE Transactions on Power Systems}, vol.~30, no.~2, pp. 621--631,
  March 2015.

\bibitem{miles_julia2015}
M.~Lubin and I.~Dunning, ``Computing in operations research using julia,''
  \emph{INFORMS J. on Comp.}, vol.~27, no.~2, pp. 238--248, 2015.

\bibitem{1519725}
Y.~V. Makarov, V.~I. Reshetov, A.~Stroev, and I.~Voropai, ``Blackout prevention
  in the united states, europe, and russia,'' \emph{Proceedings of the IEEE},
  vol.~93, no.~11, pp. 1942--1955, Nov 2005.

\bibitem{7268773}
Y.~Dvorkin, M.~Lubin, S.~Backhaus, and M.~Chertkov, ``Uncertainty sets for wind
  power generation,'' \emph{IEEE Transactions on Power Systems}, vol.~31,
  no.~4, pp. 3326--3327, July 2016.

\end{thebibliography}

\end{document}